\documentclass{ws-ijmpc}

\begin{document}

\markboth{R. Wang et al.,} {Effects of Langmuir kinetics on two-lane
TASEP}

\catchline{}{}{}{}{}

\title{EFFECTS OF LANGMUIR KINETICS OF TWO-LANE TOTALLY ASYMMETRIC EXCLUSION
PROCESSES IN PROTEIN TRAFFIC}

\author{RUILI WANG\footnote{Corresponding author.}}
\address{Institute of Information Sciences and Technology, Massey University, New Zealand \\r.wang@massey.ac.nz}

\author{RUI JIANG}
\address{School of Engineering Science, University of Science
and Technology of China, Hefei 230026, China \\ rjiang@ustc.edu.cn}

\author{MINGZHE LIU}
\address{Institute of Information Sciences and Technology, Massey University, New Zealand \\m.z.liu@massey.ac.nz}

\author{JIMING LIU}
\address{School of Computer Science, University of Windsor, Windsor, Ontario, Canada \\ jiming@uwindsor.ca}

\author{QING-SONG WU}
\address{School of Engineering Science, University of Science
and Technology of China, Hefei 230026, China \\
qswu@ustc.edu.cn}

\maketitle

\begin{history}
\received{Day Month Year} \revised{Day Month Year}
\end{history}

\begin{abstract}
In this paper, we study a two-lane totally asymmetric simple
exclusion process (TASEP) coupled with random attachment and
detachment of particles (Langmuir kinetics) in both lanes under open
boundary conditions. Our model can describe the directed motion of
molecular motors, attachment and detachment of motors, and free
inter-lane transition of motors between filaments. In this paper, we
focus on some finite-size effects of the system because normally the
sizes of most real systems are finite and small (e.g., size $\leq
10,000$). A special finite-size effect of the two-lane system has
been observed, which is that the density wall moves left first and
then move towards the right with the increase of the lane-changing
rate. We called it the jumping effect. We find that increasing
attachment and detachment rates will weaken the jumping effect. We
also confirmed that when the size of the two-lane system is large
enough, the jumping effect disappears, and the two-lane system has a
similar density profile to a single-lane TASEP coupled with Langmuir
kinetics. Increasing lane-changing rates has little effect on
density and current after the density reaches maximum. Also,
lane-changing rate has no effect on density profiles of a two-lane
TASEP coupled with Langmuir kinetics at a large
attachment/detachment rate and/or a large system size. Mean-field
approximation is presented and it agrees with our Monte Carlo
simulations.

\end{abstract}

\section{Introduction}

Recently, driven diffusive systems have attracted the interests of
physicists  because the systems show surprisingly rich and complex
behavior.\cite{SCHMITTMANN90,KRU91,TRIPATHY97,LEUNG99,MONETTI02,LEE06}
As a simple model of driven diffusive systems, asymmetric simple
exclusion processes (ASEPs) have been widely studied in chemistry
and physics.\cite{DER98,SCH01} Moreover, ASEPs have been applied
successfully in biology such as gel electrophoresis,\cite{WID91}
protein synthesis,\cite{SCHUTZ97,SHA03} mRNA translation
phenomena,\cite{CHO03} motion of molecular motors along the
cytoskeletal filaments,\cite{KLU03,JIANG05} and the depolymerization
of microtubules by special enzymes\cite{KLE05} as well as vehicular
traffic.\cite{HUANG02}

A totally asymmetric simple exclusion process (TASEP) is regarded as
the minimal model of ASEPs, in which particles move along one
direction. TASEP has received much attention in modeling
intra-cellular molecular transport and related traffic
problems\cite{KLE05,MAC68,KRU02,LIP01,PAR03,NIS05,CHO05} in recent
years.

MacDonald et al.\cite{MAC68} initially used a TASEP to simulate the
dynamics of ribosomes moving along messenger RNA chains. Kruse and
Sekimoto\cite{KRU02} proposed a two-headed TASEP model to describe
molecular motor traffic. In their model, each motor consists of two
heads and each head is attached to a filament. Each filament is
assumed to be moveable. The investigation indicates that the average
relative velocity of filaments is a non-monotonic function of the
concentration of motors. The density and current profiles of motors
with different types of microtubule tracks (e.g., cylindrical
geometry and radial geometry) have been investigated by Lipowsky et
al.\cite{LIP01}

An extension of single-lane TASEP, incorporating Langmuir kinetics
(LK, the attachment and detachment of particles), has been presented
by Parmeggiani et al.\cite{PAR03} Their model is also referred to as
the PFF model. Numerical results of the PFF model show that there
are unexpected stationary regimes for large but finite systems. Such
regimes are represented by phase coexistence in both low and high
density regions, which are separated by a domain wall.

A model, incorporating a TASEP, Langmuir kinetics and Brownian
ratchet mechanism,\cite{AST97} is proposed by Nishinari et
al.\cite{NIS05} to mimic the movement of the single-headed kinesin
motor, KIF1A. A novel feature in their model is that there are
\emph{three} states (strong binding, weak binding and no binding) of
a KIF1A, compared to \emph{two} states (binding or unbinding) in
previous models. Their model can capture explicitly the effects of
adenosine triphosphate (ATP) hydrolysis as well as the ratchet
mechanism which drives individual motors. The experimentally
observed single molecular properties in the low-density limit are
reproduced and a phase diagram is presented.

Most previous work on modeling molecular motor traffic deals with
single-lane systems where particles can move forward or backward, or
attach and detach to a bulk (a collection of sites in a lane except
boundaries). Obviously, the description of the traffic of motors
would be more realistic if multi-lane asymmetric exclusion processes
can be considered. Experimental observations\cite{HOW01} have found
that motor protein kinesins can move along parallel protofilaments
of microtubules and they can jump between these protofilaments
without restraint.

Recently, Pronina and Kolomeisky\cite{PRO04} proposed a two-lane
traffic model to simulate a TASEP with symmetric lane-changing
rules between two lanes, but without Langmuir kinetics. The
computational results suggest that lane-changing rates have a strong
effect on the steady-state properties of the system. In particular,
the particle current of each lane will decrease and particle
densities will increase with the increase of particle coupling.
Pronina and Kolomeisky then extended their model to a more general
case where asymmetric coupling is applied.\cite{PRO06} It is found
that asymmetric coupling between lanes leads to a very complex phase
diagram, quite different from symmetric coupling. There are seven
phases in the TASEP with asymmetric lane-changing rates, in contrast
to three phases found in the system with symmetric coupling. In
addition, a new maximal-current phase with a domain wall in the
intermediate coupling is reported in Ref.\cite{PRO06}

Effects of synchronization of kinks (i.e., domain walls) in a
two-lane TASEP without Langmuir kinetics is investigated by Mitsudo
and  Hayakawa.\cite{MIT05} The asymmetric lane-changing rate between
lanes is used. Moreover, different injection and ejection rates of
particles at the boundaries of two lanes are also considered. The
positions of kinks are reported to be synchronized, though the
number of particles may be different on these two lanes.

More recently, Jiang et al.\cite{JIANG07} introduced Langmuir
kinetics into one lane of a two-lane system. This has shown that
synchronization of shocks on both lanes occurs when the
lane-changing rate exceeds a threshold. A boundary layer is also
observed at the left boundary as a finite-size effect. In their
model, attachment and detachment of particles are assumed to occur
only on one of two lanes, not both, which is unable to realistically
describe real two-lane or multi-lane molecular motor traffic.

This paper will investigate the collective effect of attaching and
detaching particles on both lanes of a two-lane system with
symmetric inter-lane coupling. In particular, we will focus on the
finite-size effects of the system. The model described in this paper
is directly motivated by the dynamics of molecular motors, for
instance, unidirectional motion of molecular motors along
filaments,\cite{ALB94} random motor (e.g., kinesin) attachments and
detachments to filaments,\cite{OKA99} and molecular motors freely
changing to adjacent filaments.\cite{HOW01} It is expected that the
incorporated process of a TASEP, Langmuir kinetics and
lane-changing can shed light on the study of the traffic of
molecular motors and other particle traffic in biology as well as in
other areas such as vehicular traffic.

We investigate the effects of the following parameters on density
and current profiles: (i) Different lane-changing rates $\omega$. We
denote $\Omega = \omega N$, where $\Omega$ is used to represent the
number of lane-changing particles and $N$ is the length of each lane
(i.e., the number of cells in each lane); (ii) Different
combinations of attachment ($\Omega_A= \omega_A N$) and detachment
($\Omega_D= \omega_D N$) rates (for simplicity, we assume that
attachment rate $\Omega_A$ is fixed but detachment rate $\Omega_D$
varies); and (iii) Different system size $N$, i.e., the length of
each lane. Note that as we only consider symmetric lane changing,
these two lanes are equivalent. We compare results obtained from our
model when $\Omega
> 0$ and $\Omega = 0$ (where the two-lane system becomes two
separated TASEPs coupled with Langmuir kinetics). In this way, we
can more clearly distinguish the effect of different values of
$\Omega$ on a two-lane system. Finally, mean-field approximation
(MFA) is presented and used to verify our Monte Carlo simulations
(MCS).

The paper is organized as follows. In Section II, we give a
description of our two-lane TASEP model, considering attachment and
detachment of particles in both lanes. In Section III, we present
and discuss the results of our Monte Carlo simulations. In Section
IV, mean-field approximation is presented and the quantitative
agreement between the Monte Carlo simulations and the mean-field
theory is confirmed. Finally, we give our conclusions in Section V.

\section{Model}

Our model is defined in a two-lane lattice of $N \times 2$ sites,
where $N$ is the length of a lane. We assume that all particles move
from the left to the right, as shown in Fig. 1. Sites $i$ = 1 and
$i$ = $N$ define the left and right boundaries respectively, while a
set of sites $i$ = 2, ..., $N-1$ is referred to as a bulk. Particles
in the system are involved in the following processes: lane
selection, particle injection into the first site of a lane,
particle detachment from a bulk, particle movement along a bulk,
particle lane-changing between bulks, particle attachment to a bulk,
and particle ejection from the last site of a lane. Thus, we assume
that a particle that is injected to a lane and moves along the lane
may be involved in the following processes:

\begin{itemize}
\item Lane selection: A lane is randomly chosen.
\item Particle injection: If the first site of the selected lane is unoccupied, a particle will be
injected to the site with probability $\alpha$.
\item Particle detachment: A particle can leave the system with probability
$\omega_D$.
\item Particle movement: If a particle cannot leave the system, it can move
forward to the next site in the same lane provided that the next
site is empty. If the next site is not empty, the particle cannot
advance. In this case, the particle will check if it can change to
the other lane.
\item Particle lane-changing: The particle may change to the corresponding site on the other
lane with probability $\omega$ if the corresponding site on the
other lane is unoccupied.
\item Particle attachment:  If a site
in a bulk is empty, a particle can attach to the site with
probability $\omega_A$.
\item Particle ejection: If a particle reaches the last site of a lane, it will be ejected
from the system with probability $\beta$.
\end{itemize}
\begin{figure}[!h]
\begin{center}
\includegraphics[width= 5 in, height=3 in]{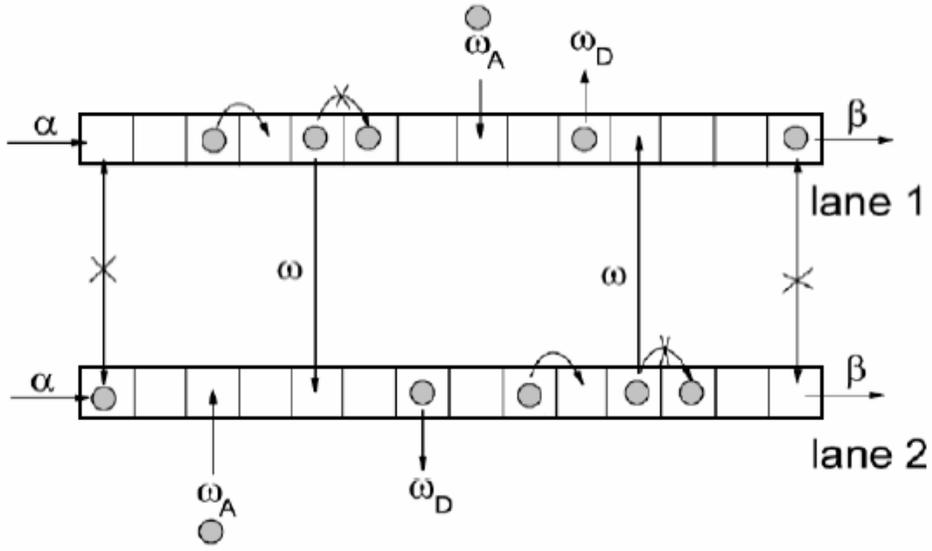}
\caption{The schematic representation of a two-lane TASEP with
symmetric inter-lane coupling, and the attachment and detachment of
particles on both lanes.}
\end{center}
\end{figure}
The above processes can be described using the following updating
rules. As the two lanes are equivalent, they have the same updating
rules. We only list the updating rules for lane 2. An occupation
variable ($\tau_{\ell,i}$) denotes the state of the $i$th site in
the $\ell$th lane, where $\tau_{\ell,i}=1$ (or $\tau_{\ell,i}=0$)
corresponds to whether the site is occupied (or empty). Thus,
$\tau_{2,i}$ refers to the state of site $i$ on lane 2. The rules
are:

\begin{itemize}
\item Case $i=1$. (i) If $\tau_{2,1}=0$, a particle enters the system with probability
$\alpha$; (ii) If $\tau_{2,1}=1$ and $\tau_{2,2}=0$, then the
particle in the site of $(2,1)$ moves into site $(2,2)$; (iii) If
$\tau_{2,1}=1$ and $\tau_{2,2}=1$, then the particle in the site of
$(2,1)$ stays there. No lane change occurs.\footnote{The simulation
results show that they are essentially the same if lane-changing is
allowed on the first site.}
\item Case $i=N$. If $\tau_{2,N}=1$, the particle leaves the system with probability $\beta$.
No lane change occurs.\footnote{The simulation results also show
that there are no differences on density profiles whether
lane-changing is considered in the last site.}
\item Case $1<i<N$. (i) If $\tau_{2,i}=1$, the particle may leave the system with probability $\omega_D$;
If it cannot leave the system, it moves into site $(2,i+1)$ provided
$\tau_{2,i+1}=0$. If the particle cannot advance, it may change to
lane 1 with probability $\omega$ if $\tau_{1,i}=0$; (ii) If
$\tau_{2,i}=0$, a particle enters the system with probability
$\omega_A$.
\end{itemize}

These updating rules for both lanes are illustrated in Fig. 1. For
simplicity, we use a symmetric lane-changing rule. It will be the
next step of our work to investigate the effects of asymmetric
lane-changing rates on the dynamics of interacting particles.

Normally, there are three types of updating schemes:\cite{LIA05} (i)
In parallel. Updating rules are synchronously applied to all sites.
A typical application is in vehicular traffic flow; (ii) In ordered
sequence. Positions of particles are updated in an ordered
sequential manner, e.g., from one end to the other end of a
one-dimensional system; (iii) In random. A site is randomly chosen
and it is updated according to updating rules. Here we use a
randomly sequential updating scheme, i.e., the third scheme, which
has been widely used in the simulations of molecular motor
traffic.\cite{KRU02,LIP01,PAR03,NIS05,PRO04,PRO06,JIANG07}

\begin{figure}[!h]
\begin{center}
\includegraphics[width= 5 in, height=3 in]{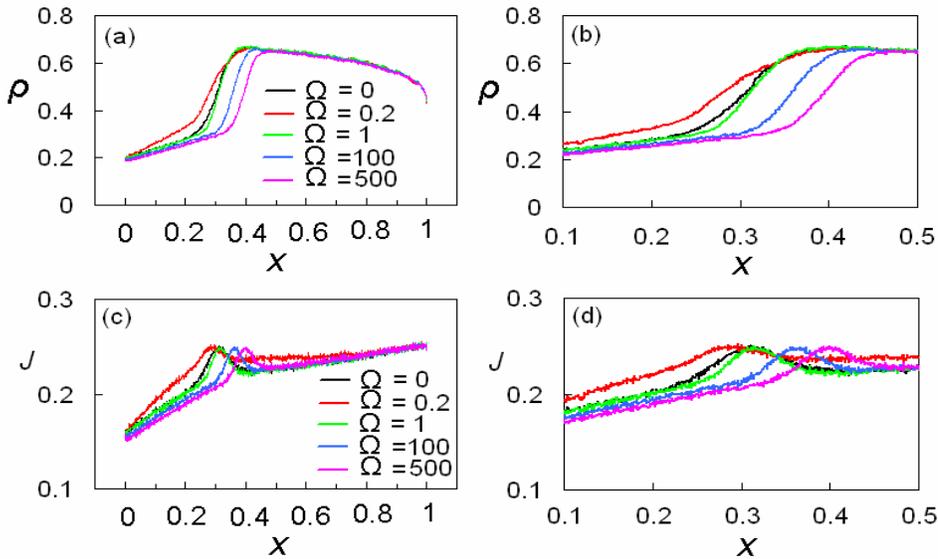}
\caption{(Color online) (a) and (b) show average density $\rho(x)$;
(c) and (d) show current $J(x)$ for different values of $\Omega$.
The system parameters are set to: $N=1000$, $\Omega_A=0.3$,
$\Omega_D=0.1$, $\alpha=0.2$ and $\beta=0.6$. (b) and (d) are the
locally enlarged figures of (a) and (c) respectively.}
\end{center}
\end{figure}

\section{Monte Carlo Simulations}

In this section, the results of Monte Carlo simulations are
presented. The average current ($J_i$) is given by: $J_i$ =
$\rho_i$(1 - $\rho_{i+1})$. The parameters are set to $N=1,000$,
$\alpha=0.2$, $\beta=0.6$, $\Omega_D=0.2$, $K=3$ and
$\Omega_A$=$K$*$\Omega_D$. In simulations, stationary profiles are
obtained by averaging $10^5$ samples at each site. The sampling
time interval is $10N$. The first $10^5N$ time steps are discarded
to let the transient time out. For the special case when $\Omega=0$,
the two-lane process is reduced to two separated processes: two
one-lane TASEP with Langmuir kinetics. Features of a one-lane TASEP
with Langmuir kinetics have been well studied in Ref.\cite{PAR03}

Fig. 2 shows average density $\rho(x)$ and current $J(x)$ of one
lane of a two-lane system for different values of $\Omega$. For the
special case $\Omega = 0$, the two-lane system reduces to two
separated single-lane systems. When $\Omega$ increases from  0 to 1,
we may expect the domain wall always moves in one direction (either
to the left or right). However, an unexpected phenomenon has been
observed in our simulations: the domain wall first moves to the left
slightly, and then moves towards the right with the increase of
$\Omega$ (see Figs. 2(a) and (b)). This process is more clearly
captured in Figs. 3(a) and (b), where the domain walls for $\Omega
\leq 0.4 $ are shown. One can see that the domain wall moves to the
left first when $\Omega$ increases from 0 to 0.2, and then moves
towards the right when $\Omega > 0.2$. We refer to this feature as
the \emph{jumping effect} to describe the domain wall (as well as
current curve) jump to the left and then to the right caused by the
particles jumping between the two lanes. We also find that there is
a critical value,  $\Omega_c \approx 0.2$, to distinguish the
movement direction of the domain wall under certain conditions. A
similar effect occurs in Ref.\cite{JIANG07}

\begin{figure}[!h]
\begin{center}
\includegraphics[width= 5 in, height=2 in]{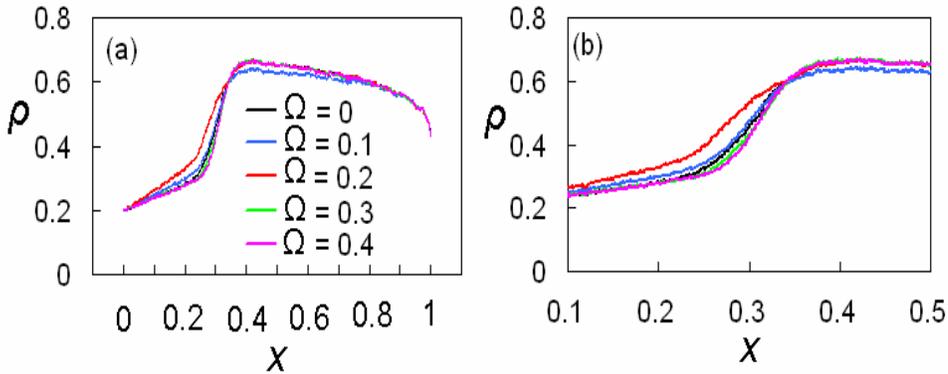}
\caption{(Color online) Average density $\rho(x)$ for small values
of $\Omega$. The system parameters are: $N=1000$, $\Omega_A=0.3$,
$\Omega_D=0.1$, $\alpha=0.2$ and $\beta=0.6$.}
\end{center}
\end{figure}

When $\Omega_A$ and $\Omega_D$ increase proportionally, the jumping
effect becomes weaker (comparing Figs. 2(a) and (b)with Figs. 4(a)
and (b)). This implies that the jumping effect is also influenced by
attachment and detachment rates.

\begin{figure}[!h]
\begin{center}
\includegraphics[width= 5 in, height=3 in]{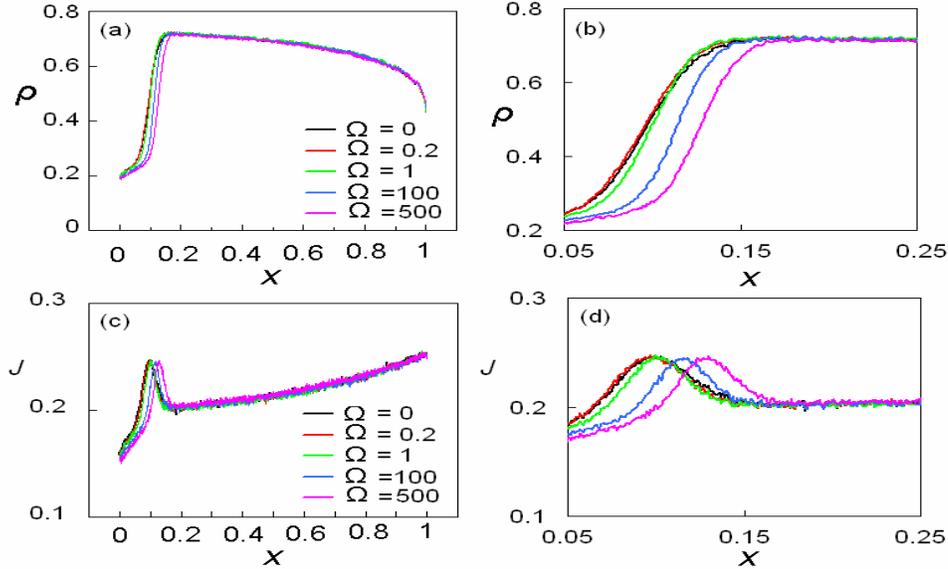}
\caption{(Color online) (a) Average density $\rho(x)$; Current
$J(x)$ for different values of $\Omega$. The system parameters are:
$N=1,000$, $\Omega_A=0.6$, $\Omega_D=0.2$, $\alpha=0.2$ and
$\beta=0.6$. (b) and (d) are locally enlarged figures of (a) and (c)
respectively.}
\end{center}
\end{figure}

The jumping effect is also observed in a larger system, e.g.,
$N=10,000$, (see Fig. 5) when $\Omega$ is small. It can be seen that
the domain wall moves left when $\Omega \leq 0.3$. When $\Omega
>0.3$, the domain wall moves right. For instance, there is an
obvious movement towards right when $\Omega=0.4$ (see Fig. 5(b)).
The critical value $\Omega_c$ of $\Omega$ varies with a system size,
i.e., $\Omega_c$ changes from 0.2 to 0.3 when the system size
increases from 1,000 to 10,000.

We have also confirmed that the jumping effect in a two-lane
homogeneous TASEP coupled with Langmuir kinetics is a kind of
finite-size effect. When the system size is very large, e.g.,
$N=100,000$, in our simulations, the jumping effect almost
disappears.

We argue that we need to study and examine finite-size effects, like
the jumping effect because the size of a real system is normally not
very large. For example, kinesin protein motors are responsible for
long-distance transport in a cell. The length that a kinesin protein
motor travels from its origin to its destination is normally about
100 successive steps on microtubules and the step size of kinesins
is about 8nm.\cite{BLOCK00} This indicates that a system size below
1,000 is enough to realistically describe the movements of molecular
motors. Even for a large system, a size $N \leq 10,000$ is normally
adequate for simulation. Thus, a conclusion obtained based on the
assumption of an infinite size may not be applicable to any
real system.

Comparing Figs. 2(a) and (b) with Figs. 4(a) and (b), it can be seen
that the slopes of domain walls increase with the increase of
$\Omega_A$ and $\Omega_D$ proportionally. This is because when the
difference between $\Omega_A$ and $\Omega_D$ enlarges, more
particles will attach to these two lanes rather than detach from
them.

Increasing the value of $\Omega$ means more particles have chances
to change to the other lane if they cannot move forward along the
current lane. Upon further increasing $\Omega$, it is found that the
density decreases, and the locations of shocks shift right slightly
(see Figs. 2(a) and 3(a)).

When $x > x_{max}^{den}$ ($x_{max}^{den}$ denotes the position where
the density reaches maximum), the differences between density
profiles are not obvious (see Figs. 2 and 4). This suggests that
lane-changing rate $\Omega$ does not have any obvious effect on the
system properties (e.g., average density and current profiles) after
the maximum density is reached.

In Figs. 2(c) and (d), and 4(c) and (d), current profiles are
illustrated. It is found that (i) The current decreases with the
increase of $\Omega$ until it approaches the maximum current
($J_{max}$). When $\rho_{i+1}$ is approximately equal to $\rho_i$,
the current $J_i \approx$ $\rho_i$(1 - $\rho_i)$. Thus, the maximum
current $J_{max} \approx$ 0.25. We denote by $x_{max}^{cur}$
the position where the current reaches the maximum; (ii) When
$x_{max}^{cur} < x < x_{max}^{den}$, the current increases when
lane-changing rate $\Omega$ increases; (iii) When $x >
x_{max}^{den}$, the current profiles are almost the same for
different $\Omega$.

\begin{figure}[!h]
\begin{center}
\includegraphics[width= 5 in, height=2 in]{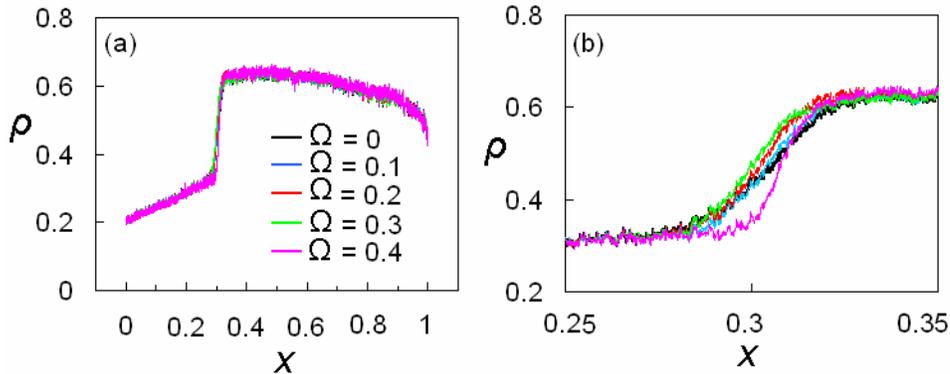}
\caption{(Color online) Average density $\rho(x)$ for small values
of $\Omega$ in a large system. The system parameters are: $N=10000$,
$\Omega_A=0.3$, $\Omega_D=0.1$, $\alpha=0.2$ and $\beta=0.6$.}
\end{center}
\end{figure}

\begin{figure}[!h]
\begin{center}
\includegraphics[width= 5 in, height=2 in]{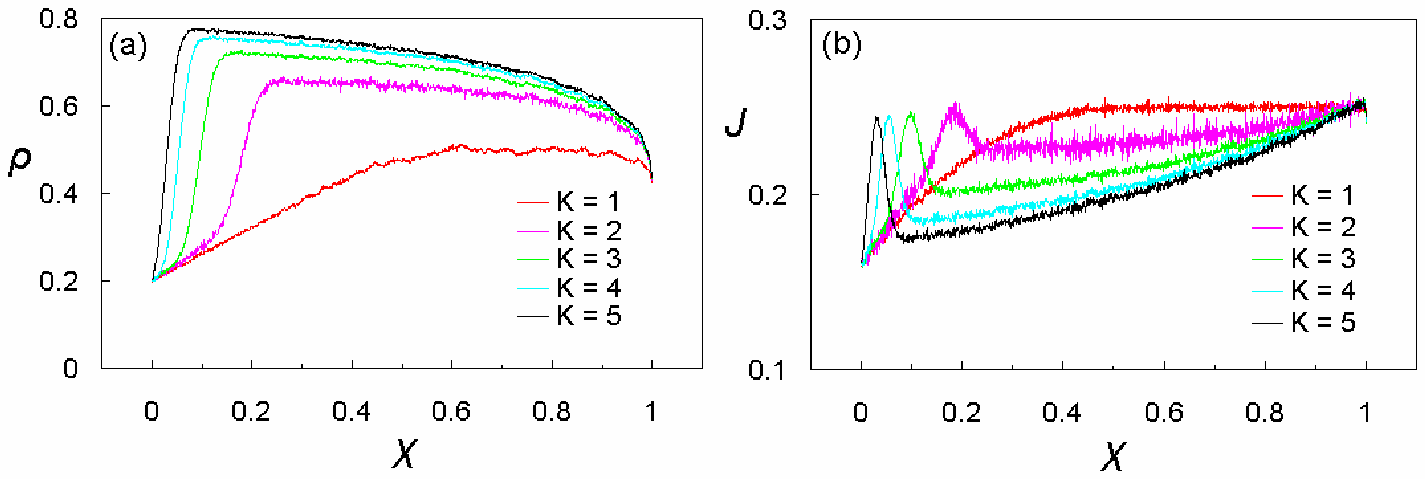}
\caption{(Color online) Average density $\rho(x)$ for different
values of $K$. The system parameters are: $N=1000$, $\Omega_A=0.3$,
$\Omega_D=0.1$, $\alpha=0.2$ and $\beta=0.6$.}
\end{center}
\end{figure}

We next investigate the effect of the values of $K$
($K=\Omega_A$/$\Omega_D$) on average density and current profiles.
We consider the situation in which $\Omega_A$ is fixed but
$\Omega_D$ varies. Note that the increase of $K$ means the decrease
of $\Omega_D$. It can be seen that average density increases with
the increase of the value of $K$ (see Fig. 6(a)). Moreover, the
shock moves left and the amplitude increases. In other words, the
width of the transition region decreases. This may be explained as
follows: when the value of $K$ increases, the value of $\Omega_D$
decreases proportionally, so that the opportunities for particles to
detach from the bulk become smaller. In this case, more particles
will remain in the bulk. This leads to an increase of average
density.

In Fig. 6(b), the current first increases with the increase of the
value of $K$ till it reaches the maximum value, and then the current
decreases with the increase of the value of $K$. In other words, the
position with the maximum current moves left and the amplitude
increases (see Fig. 6(b)) upon increasing $K$. The current curves
can be obtained from equation $J_i$ = $\rho_i$(1 - $\rho_{i+1})$, it
thus can be seen that the maximum current $J_{max}$ =0.25 when
densities of two adjacent sites are equal to 0.5. When density is
higher or lower than 0.5,  the corresponding current is lower than
0.25.

\begin{figure}[!h]
\begin{center}
\includegraphics[width= 5 in, height=2 in]{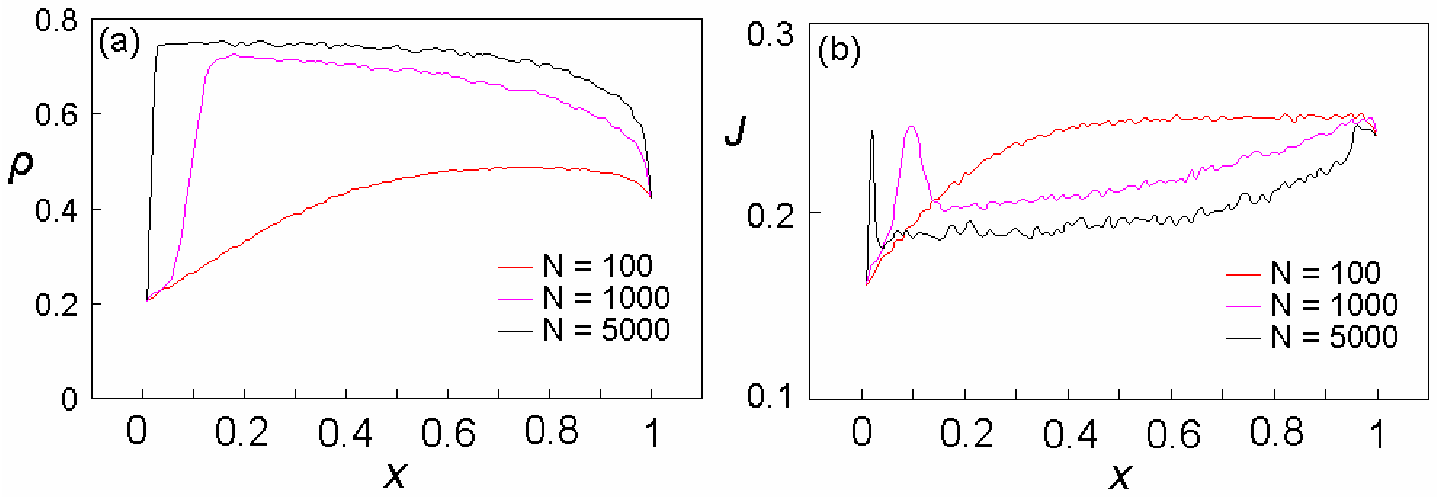}
\caption{(Color online) (a) Average density $\rho(x)$; (b) current
$J(x)$ for different values of the system size. The system
parameters are: $\Omega=0.2$, $\Omega_A=0.6$, $\Omega_D=0.2$,
$\alpha=0.2$ and $\beta=0.6$.}
\end{center}
\end{figure}

Fig. 7 shows the density and current profiles for different system
sizes. It is found that the width of the transition region decreases
as the number of sites is increased (see Fig. 7(a)). Our simulation
suggests the boundary layer shows a finite-size effect and will
disappear in the limit of an infinite system. The finite-size effect
shown in Fig. 7(a) is consistent with that in
Refs.\cite{PAR03,JIANG07} In Fig. 7(b) one observes the amplitude in
the left boundary enhances when the system size increases.

\section{Mean-field Approximation}

In this section, a mean-field theory is developed. The occupation
variable ($\tau_{\ell,i}$) defined in section II is also used. The
corresponding equation for the evolution of particle densities
$\langle\tau_{\ell,i}\rangle$ in a bulk (i.e., $1<i<N)$ can be
written as

\begin{eqnarray}
\lefteqn{
\frac{d\langle\tau_{1,i}\rangle}{dt}=\langle\tau_{1,i-1}(1-\tau_{1,i})\rangle-\langle\tau_{1,i}(1-\tau_{1,i+1})\rangle{}
} \nonumber\\
& &
{}+\omega\langle\tau_{2,i}\tau_{2,i+1}(1-\tau_{1,i})\rangle-\omega\langle\tau_{1,i}\tau_{1,i+1}(1-\tau_{2,i})\rangle{}
\nonumber\\
&
&+\omega_A\langle1-\tau_{1,i}\rangle-\omega_D\langle\tau_{1,i}\rangle,
\end{eqnarray}

\begin{eqnarray}
\lefteqn{
\frac{d\langle\tau_{2,i}\rangle}{dt}=\langle\tau_{2,i-1}(1-\tau_{2,i})\rangle-\langle\tau_{2,i}(1-\tau_{2,i+1})\rangle{}
} \nonumber\\
& & {}
+\omega\langle\tau_{1,i}\tau_{1,i+1}(1-\tau_{2,i})\rangle-\omega\langle\tau_{2,i}\tau_{2,i+1}(1-\tau_{1,i})\rangle{}
\nonumber\\
&
&+\omega_A\langle1-\tau_{2,i}\rangle-\omega_D\langle\tau_{2,i}\rangle,
\end{eqnarray}
where $\langle \cdots \rangle$ denotes a statistical average. The
term $\omega\langle\tau_{1,i}\tau_{1,i+1}(1-\tau_{2,i})\rangle$ is
the average current from site $i$ in lane 1 to site $i$ in lane 2,
while $\omega\langle\tau_{2,i}\tau_{2,i+1}(1-\tau_{1,i})\rangle$ is
the average current from site $i$ in lane 2 to site $i$ in lane 1.
The first two terms in Eqs. (1) and (2) correspond to particle
movement; the middle two terms correspond to particle lane changing,
and the last two terms correspond to attachment and detachment of
particles. At the boundaries, the densities evolve as
\begin{equation}
\label{m3} \frac{d\langle \tau_{\ell,1}\rangle}{dt}=\alpha\langle
1-\tau_{\ell,1}\rangle-\langle
\tau_{\ell,1}(1-\tau_{\ell,2})\rangle,
\end{equation}
\begin{equation}
\label{m4} \frac{d\langle \tau_{\ell,N}\rangle}{dt}=\langle
\tau_{\ell,N-1}(1-\tau_{\ell,N})\rangle -\beta\langle
\tau_{\ell,N}\rangle,
\end{equation}
where $\ell$ can be 1 or 2. It can be seen that Eqs. (1) and (2) are
equivalent when we use the same lane-changing rate between two
lanes. Therefore, we only focus on the mean-field approximation in
lane 2 (i.e., Eq. (2)). When $N$ is large, i.e., $N\gg 1$, we can
transfer the coarse-grained stationary-state Eq. (2) to continuum
mean-field approximation. First, we factorize correlation functions
by replacing $\langle \tau_{\ell,i}\rangle$ and $\langle
\tau_{\ell,i+1} \rangle$ with $\rho_{\ell,i}$ and $\rho_{\ell,i+1}$,
then we set
\begin{equation}
\label{m5} \rho_{\ell,i\pm 1}=\rho(x)\pm \frac{1}{N}\frac{\partial
\rho}{\partial x}+\frac{1}{2N^2}\frac{\partial^2 \rho}{\partial
 x^2}+ O(\frac{1}{N^3}).
\end{equation}

Substituting (\ref{m5}) into Eq. (2), we obtain
\begin{eqnarray}
\label{m6} \lefteqn{ \frac{\partial \rho_2}{\partial
t'}=(2\rho_2-1)\frac{\partial \rho_2}{\partial
x}+\Omega_A(1-\rho_2)-\Omega_D \rho_2{} } \nonumber\\
& & -\Omega (1-\rho_1)\rho_2^2+\Omega \rho_1^2(1-\rho_2){},
\end{eqnarray}
where $t'=t/N$, $\Omega=\omega N$, $\Omega_A=\omega_A N$ and
$\Omega_D=\omega_D N$. The term
$\langle\tau_{1,i}\tau_{1,i+1}(1-\tau_{2,i})\rangle$ is replaced by
$\rho_{1,i}\rho_{1,i+1}(1-\rho_{2,i})$, and $\rho_{1,i+1}$ is
approximated as $\rho_{1,i} $; similarly, the term
$\langle\tau_{2,i}\tau_{2,i+1}(1-\tau_{1,i})\rangle$ is replaced by
$\rho_{2,i}\rho_{2,i+1}(1-\rho_{1,i})$, and $\rho_{2,i+1}$ is
approximated as $\rho_{2,i} $. The boundary conditions become
$\rho_{\ell}(x=0)=\alpha$ and $\rho_{\ell}(x=1)=(1-\beta)$. As
lane-changing is symmetric and two lanes are homogeneous, we have
$\rho_2=\rho_1=\rho$ (our Monte Carlo simulations show that
densities in lanes 1 and 2 are the same). Thus, Eq. (6) reduces to
\begin{eqnarray}
\label{m7} \frac{\partial \rho}{\partial t'}=(2\rho-1)\frac{\partial
\rho}{\partial x}+\Omega_A(1-\rho)-\Omega_D \rho.
\end{eqnarray}

In the limit of $t \rightarrow \infty$, the system reaches a
stationary state with $\frac{\partial \rho}{\partial t'}$ = 0, Eq.
(7) simplifies into

\begin{equation}
\label{m8} (2\rho-1)\frac{\partial \rho}{\partial
x}+\Omega_A(1-\rho)-\Omega_D \rho=0.
\end{equation}

Eq. (\ref{m8}) is a first order differential equation and has been
solved in Ref.\cite{EVANS03} The low-density profiles in the bulk
can be obtained by integrating the equation from the left boundary
$(\rho(0)=\alpha)$ to a density $\rho_l$:
\begin{equation}
\label{m9}
x=\frac{1}{\Omega_D(K+1)}[2(\rho_l-\alpha)+\frac{K-1}{K+1}\ln|\frac{K-(K+1)\rho_l}{K-(K+1)\alpha}|].
\end{equation}
Similarly, we can also integrate the equation from the right
boundary $(\rho(1)=1-\beta)$ to a density $\rho_r$:
\begin{equation}
\label{m10}
1-x=\frac{1}{\Omega_D(K+1)}[2(1-\beta-\rho_r)+\frac{K-1}{K+1}\ln|\frac{K-(K+1)(1-\beta)}{K-(K+1)\rho_r}|],
\end{equation}
where $K=\Omega_A/\Omega_D$. The overall density profile across the
system can be described by determining a shock position in these two
profiles. In Fig. 8, we compare the results of mean-field
approximation (MFA) with the results of Monte Carlo simulations
(MCS). We find that the results of Monte Carlo simulations are in
agreement with that of mean-field approximation. We note that
numerical results of MCS in Figs. 8(a) and (b) are almost the same
as Figs. 2 and 5 in Ref.,\cite{EVANS03} which also suggests that
lane-changing rate $\Omega$ has no effect on density profiles of a
two-lane TASEP coupled with Langmuir kinetics at a large
attachment/detachment rate and/or a large system size.

\begin{figure}[!h]
\begin{center}
\includegraphics[width= 5 in, height=3 in]{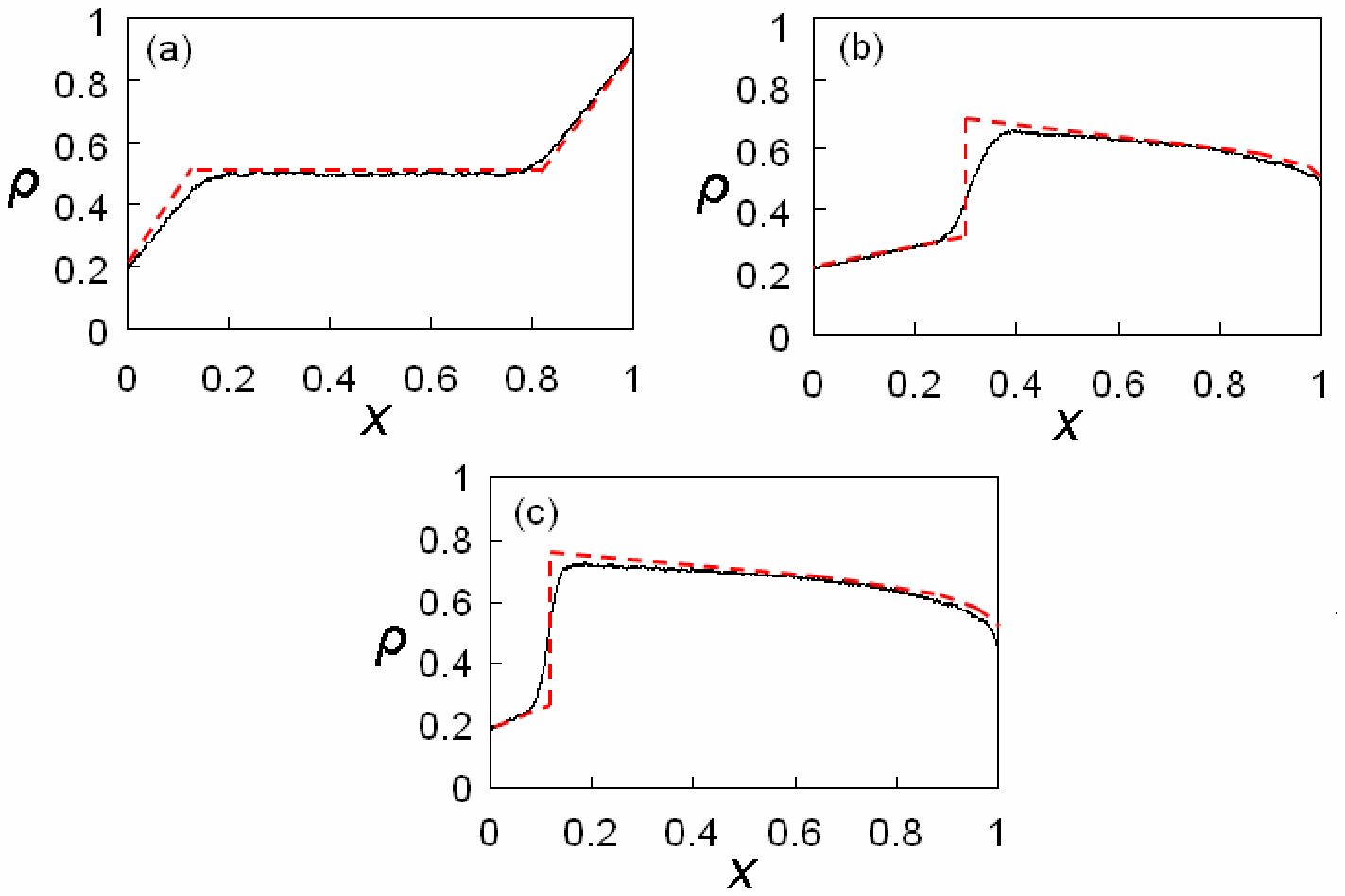}
\caption{Density profiles from Monte Carlo simulations (black line)
and mean-field approximation (red dashed line). (a) $\alpha=0.2,
\beta=0.1, k=1, \Omega_D=0.1$ and $\Omega=0.2$; (b) $\alpha=0.2,
\beta=0.6, k=3, \Omega_D=0.1$ and $\Omega=2$; (c) $\alpha=0.2,
\beta=0.6, k=3, \Omega_D=0.2$ and $\Omega=100$. The system size
$N=1,000$.}
\end{center}
\end{figure}

\section{Conclusion}

In this paper, two-lane totally asymmetric exclusion processes
coupled with Langmuir kinetics on both lanes are studied using Monte
Carlo simulations (MCS) and mean-field approximation (MFA). The
results of Monte Carlo simulations agree with the results of
mean-field approximation. The model is directly inspired by the
experimentally observed movement of molecular motors which include
motor advancing along filaments, random attachment and detachment,
and free jumping between filaments.

The system has mainly demonstrated the following complex behavior on
two lanes.
\begin{itemize}
\item
The \emph{jumping} effect is observed, that is, the domain wall
first moves left slightly, and then move towards the right with the
increase of the lane-changing rate. This effect is a finite-size
effect as it is not observed when the system is large,
e.g., $N=100,000$. On the other hand, increasing attachment and
detachment rates proportionally will lead to the \emph{jumping}
effect becoming weaker.
\item
After densities reach maximum, the increase of the lane-changing
rate has little effect on the system properties such as average
density and current.
\item
When $\Omega_A$ (attachment rate) is fixed, it is found that average
density and current increase and then the current lowers upon
decreasing the value of $\Omega_D$ (detachment rate).
\item
When system size $N$ is increased, the amplitude of the shock
increases and average density increases.
\item
Lane-changing rate $\Omega$ has almost no effect on density profiles
in a two-lane TASEP coupled with Langmuir kinetics at a large
attachment/detachment rate and/or a large system size.
\end{itemize}

\section*{Acknowledgements}
R. Wang acknowledges the support of the ASIA:NZ Foundation Higher
Education Exchange Program (2005), Massey University Research Fund
(2005), and Massey University International Visitor Research Fund
(2007). R. Jiang acknowledges the support of National Basic Research
Program of China (2006CB 705500), the National Natural Science
Foundation of China (NNSFC) under Key Project No. 10532060, Project
Nos. 10404025, 10672160 and 70601026, and the CAS Special
Foundation. We are grateful to Denise Newth for proofreading this
manuscript.

\end{document}